\documentclass[pdflatex,sn-mathphys-num]{sn-jnl}% Math and Physical Sciences Numbered Reference Style
\usepackage{graphicx}%
\usepackage{multirow}%
\usepackage{amsmath,amssymb,amsfonts}%
\usepackage{amsthm}%
\usepackage{mathrsfs}%
\usepackage[title]{appendix}%
\usepackage{xcolor}%
\usepackage{textcomp}%
\usepackage{manyfoot}%
\usepackage{booktabs}%
\usepackage{algorithm}%
\usepackage{algorithmicx}%
\usepackage{algpseudocode}%
\usepackage{listings}%
\usepackage{pythonhighlight}
\usepackage{graphicx}
\usepackage{booktabs}
\usepackage{subcaption} % Required for subfigures
\usepackage{array}      % For table formatting
\usepackage{booktabs}  % For \toprule, \midrule, \bottomrule
\usepackage{float}     % For [H] option in table
\usepackage{siunitx}   % For centered numbers
\usepackage[T1]{fontenc}
\usepackage[utf8]{inputenc}
\usepackage{threeparttable} % in preamble
\theoremstyle{thmstyleone}%
\theoremstyle{thmstyletwo}%

\theoremstyle{thmstylethree}%

\begin{document}

\title[Article Title]{CT-Less Attenuation Correction Using Multiview Ensemble Conditional Diffusion Model on High-Resolution Uncorrected PET Images}


\author*[1,2]{\fnm{Alexandre} \sur{St-Georges}}\email{Alexandre.St-Georges@USherbrooke.ca}

\author[1,2]{\fnm{Gabriel} \sur{Richard}}

\author[1,2,3]{\fnm{Maxime} \sur{Toussaint}}

\author[4]{\fnm{Christian} \sur{Thibaudeau}}

\author[4]{\fnm{Etienne} \sur{Auger}}

\author[5]{\fnm{Étienne} \sur{Croteau}}

\author[5]{\fnm{Stephen} \sur{Cunnane}}

\author[1,2,4]{\fnm{Roger} \sur{Lecomte}}

\author*[2,6]{\fnm{Jean-Baptiste} \sur{Michaud}}\email{Jean-Baptiste.Michaud@USherbrooke.ca}

\affil*[1]{\orgdiv{Department of Medical Imaging and Radiation Sciences}, \orgname{Université de Sherbrooke}, \orgaddress{\city{Sherbrooke}, \state{QC}, \country{Canada}}}

\affil[2]{\orgdiv{Sherbrooke Molecular Imaging Center}, \orgname{CRCHUS}, \orgaddress{\city{Sherbrooke}, \state{QC}, \country{Canada}}}

\affil[3]{\orgdiv{Nantes Université} \orgname{Laboratoire CRCI2NA, INSERM, CNRS,} \orgaddress{\city{Nantes}, \country{France}}}

\affil[4]{\orgdiv{} \orgname{Imaging Research and Technology (IR\&T) Inc.,} \orgaddress{\street{} \city{Sherbrooke}, \postcode{} \state{Quebec}, \country{Canada}}}

\affil[5]{\orgdiv{Research Center on Aging, Department of Medicine}, \orgname{Université de Sherbrooke}, \orgaddress{\city{Sherbrooke}, \state{QC}, \country{Canada}}}

\affil[6]{\orgdiv{Department of Electrical Engineering and Computer Engineering}, \orgname{Université de Sherbrooke}, \orgaddress{\city{Sherbrooke}, \state{QC}, \country{Canada}}}


\abstract{Accurate quantification in positron emission tomography (PET) is essential for accurate diagnostic results and effective treatment tracking. A major issue encountered in PET imaging is attenuation. Attenuation refers to the diminution of photon detected as they traverse biological tissues before reaching detectors. When such corrections are absent or inadequate, this signal degradation can introduce inaccurate quantification, making it difficult to differentiate benign from malignant conditions, and can potentially lead to misdiagnosis. Typically, this correction is done with co-computed Computed Tomography (CT) imaging to obtain structural data for calculating photon attenuation across the body. However, this methodology subjects patients to extra ionizing radiation exposure, suffers from potential spatial misregistration between PET/CT imaging sequences, and demands costly equipment infrastructure. Emerging advances in neural network architectures present an alternative approach via synthetic CT image synthesis. Our investigation reveals that Conditional Denoising Diffusion Probabilistic Models (DDPMs) can generate high quality CT images from non attenuation corrected PET images in order to correct attenuation. By utilizing all three orthogonal views from non-attenuation-corrected PET images, the DDPM approach combined with ensemble voting generates higher quality pseudo-CT images with reduced artifacts and improved slice-to-slice consistency. Results from a study of 159 head scans acquired with the Siemens Biograph Vision PET/CT scanner demonstrate both qualitative and quantitative improvements in pseudo-CT generation. The method achieved a mean absolute error of 32 $\pm$ 10.4 HU on the CT images and an average error of (1.48 $\pm$ 0.68)\% across all regions of interest when comparing PET images reconstructed using the attenuation map of the generated pseudo-CT versus the true CT.}

\keywords{Deep learning, Diffusion model, attenuation correction, pseudo-CT, image generation, brain PET, CT-less attenuation correction}



\maketitle
\vspace{-0.3cm}
\section{Background}\label{sec1}
In Positron Emission Tomography (PET), attenuation refers to the loss of photons resulting from photoelectric absorption and Compton scattering as photons interact with electrons in the body. This results in reduced signal intensity, especially in deeper regions and areas where tissues are denser. Such quantification inaccuracies can result in diagnostic errors or misinterpretation \cite{impact_attenuation}. Attenuation correction is critical for accurate diagnostics and is systematically performed in the clinic using various methods. This process requires an attenuation map, which estimates how much the emitted photons are absorbed or scattered as they travel through the body. The attenuation map is essentially an image in which each pixel represents the attenuation coefficient of the tissue - a value that describes how much the tissue reduces the intensity of passing photons.\\
\\
The most common way to generate this attenuation map is by using a secondary imaging modality that provides anatomical detail, such as Computed Tomography (CT) \cite{CT_correction} or Magnetic Resonance Imaging (MRI)\cite{MRI_correction}. CT provides a direct measurement of tissue attenuation, but since the X-ray energy range differs from that of PET annihilation photons, an energy conversion, from Hounsfield units to attenuation coefficients, is necessary to generate an accurate attenuation map \cite{Ritt2014}. MRI also offers anatomical information, but it measures proton density rather than the electron density mostly responsible for photon interactions. As a result, generating an attenuation map from MRI requires a more complex conversion process to estimate attenuation coefficients. Several methods can be used for this, including atlas-based approaches \cite{irm_atlas}, tissue segmentation\cite{irm_segmentation}, zero echo time imaging (ZTE)\cite{irm_zero_echo_time}, and deep learning techniques\cite{irm_deep_learning}. Another approach to this problem is to use a rotating source to directly measure the transmission through tissue \cite{bailey1998transmission}. However, this method requires long acquisition times, which increases as the spatial resolution of the scanner improves due to its statistical nature. This makes the approach more susceptible to patient movement and increases radiation exposure.\\
\\
Some scanners lack these attenuation correction methods and must therefore resort to new emerging technologies, such as deep learning \cite{performance_ai_attenuation}, \cite{Review_articles}. Considerable work has been done recently to generate pseudo-CT from Non-Attenuation-Corrected (NAC) PET images using various architectures, such as UNET \cite{correction_pct}, Generative Adversary Network (GAN) \cite{corr_GAN_pct} and more recently vision transformer \cite{vision_trans}. Other methods utilize the NAC PET to directly correct the NAC image with promising recent architectures such as the Denoising Diffusion Probabilistic Model (DDPM) \cite{diffusion_multiview}. DDPMs are appealing as they remove the blurry aspect provided by UNET models \cite{diffusion_original}, avoid mode collapse by GAN models \cite{mode_collapse} and generate higher quality samples than vision transformers, making them a great choice for pseudo-CT generation.\\
\\
In this work, we describe an improved version of the DDPM using orthogonal planes and majority voting demonstrating that it is possible to precisely correct attenuation without the need for another imaging modality and, most importantly, without additional radiation exposure and misregistration risk with an adjunct anatomical imaging modality.

\section{Methods}\label{sec2}

Ensuring the reliability of outputs from generative neural networks is a significant challenge because these systems can introduce artifacts and produce inconsistent results. To mitigate these concerns, our approach focuses on creating pseudo-CT (pCT) images from NAC PET scans, instead of directly producing attenuation-corrected PET images from a NAC PET image. This methodology ensures that only quantitative aspects of the PET data are modified while the anatomical structures remain untouched.  This also improves the interpretability for doctors about how exactly the PET image is going to be modified for attenuation by the pCT. Furthermore, to limit artifact generation, our proposed model generates the pCT images of the same from different orthogonal planes of the NAC PET image in order to detect and eliminate artifacts before they reach the screen of medical professionals.

\subsection{Model}

\begin{figure}[H]
    \centering
    \begin{subfigure}[c]{0.52\textwidth}
        \includegraphics[width=\textwidth]{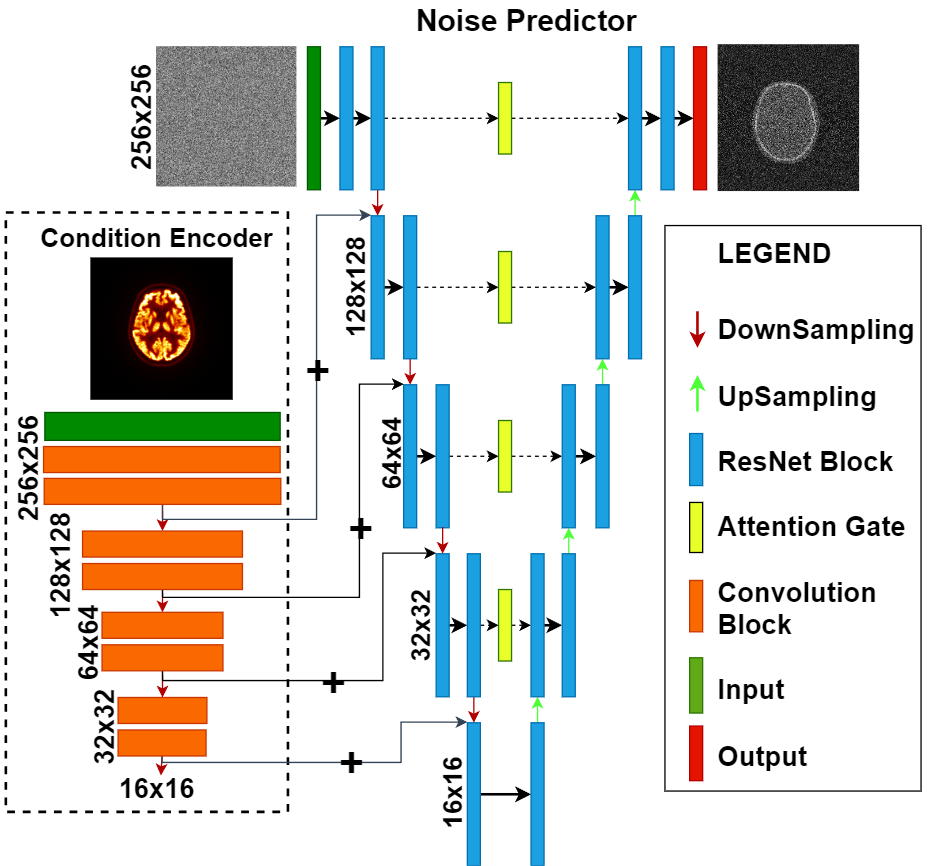}
        \caption{Architecture overview}
        \label{fig:architecture}
    \end{subfigure}
    \hfill
    \begin{subfigure}[c]{0.44\textwidth}
        \includegraphics[width=\textwidth]{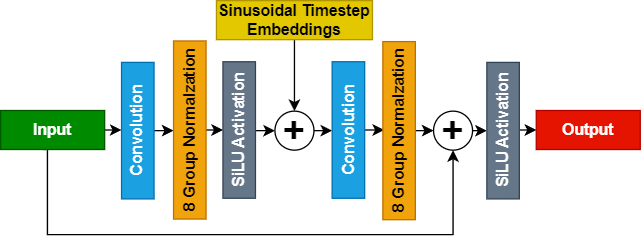}
        \caption{Architecture of the ResNet block}
        \label{fig:resnet_block}
    \end{subfigure}
    \caption{Model architecture.}
    \label{fig:sidebyside}
\end{figure}

The proposed approach uses a Denoising Diffusion Probabilistic model \cite{diffusion_original} with a UNet backbone that can be seen in Figure \ref{fig:architecture} and consists of ResNet blocks, illustrated in Figure \ref{fig:resnet_block}.
\vspace{-0.3cm}
\vspace{-0.3cm}

The proposed architecture consists of a noise predictor that removes the noise from CT images and a condition encoder that converts the PET image into a corresponding CT image. This condition encoder consists of convolution blocks that each use a 2D convolution, batch normalization and ReLU activation. The information from the condition encoder is passed to the noise predictor right after downsampling, by concatenation at stages 2 to 5. The model shown is represented for training with transverse image slices as the shape of input images is 256x256. In the case of sagittal and coronal training, this shape becomes 96x256 or 256x96 per 2D image. The attention gates come directly from Oktay \cite{attention_gates} and are used to focus on relevant spatial regions. Each stage of the model, going from up (original image shape) to down has [64, 128, 256, 512, 1024] filters for the convolutions. All downsampling and upsampling steps are performed using convolutions and transposed convolutions with a (2, 2) stride and all convolutions use a (3x3) kernel size.

\subsection{Voting}

\begin{figure}
    \centering
    \includegraphics[width=0.97\linewidth]{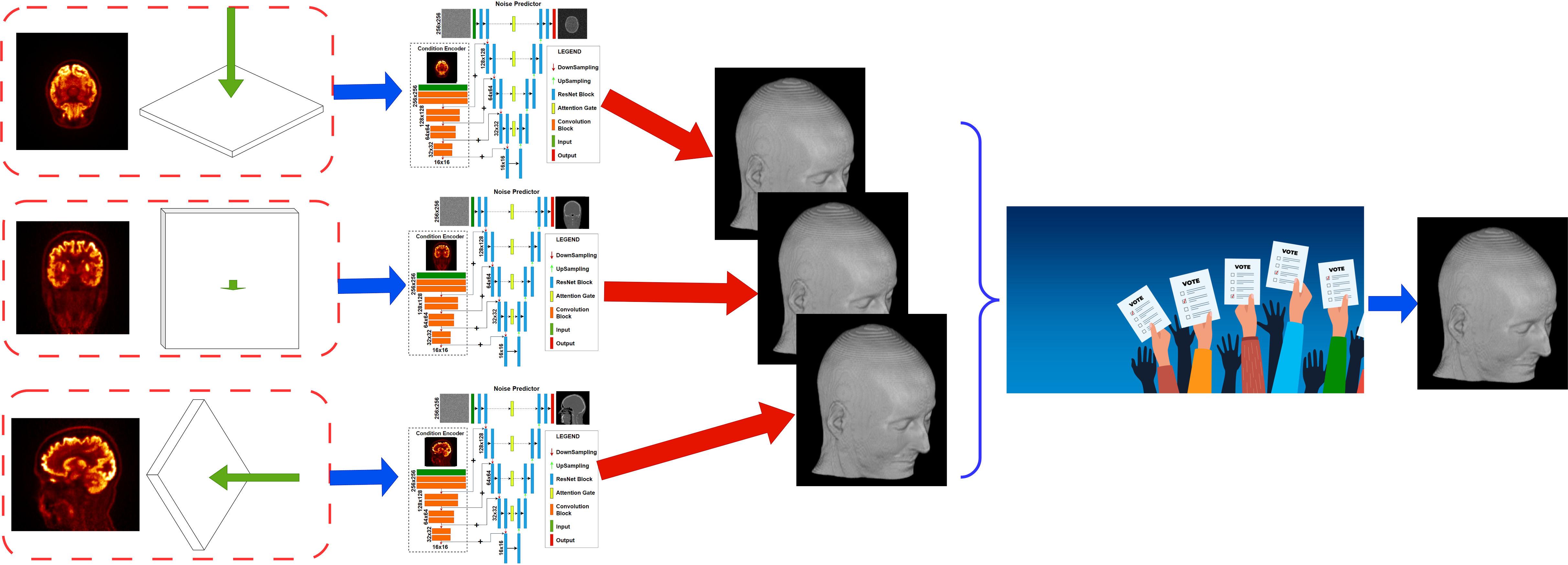}
    \caption{Voting process of three models generating attenuation-corrected PET images of the same patient from 2D image slices along three orientations.}
    \label{fig:projection}
\end{figure}

Three separate 2D models are developed, each utilizing a distinct slice orientation: transverse, sagittal, and coronal. The slices of each model are combined to create three 3D PET images of the same patient, which undergo pixel-by-pixel comparison. The voting process involves three scenarios: 1) when all three values fall within a predefined distance threshold, they are averaged together; 2) if two values meet the threshold criteria while one does not, the outlying value is excluded and the remaining two are averaged; and 3) when all three distances exceed the threshold, the two most similar values are averaged. Figure \ref{fig:projection} illustrates this process.

\subsection{Dataset}
%PET data was reconstructed with interpolated CTs and pseudo-CTs using OP algorithm provided by Siemens with a matrix size of (440x440x159) on the original PET data.
The model was trained with 159 paired PET/CT brain images from 64 men and 95 women, acquired using a \textit{Siemens Biograph Vision} scanner. Of these, 79 used [$^{18}$F]-Fluorodeoxyglucose (FDG) and 80 used [$^{11}$C]-Acetoacetate (AcAc) \cite{bentourkia}\cite{aceto_human}. Patients were on average 70 years old (range 55-81). For FDG, the tracer was injected 30 minutes before scanning, and for AcAc, immediately before. Scan times were 30 minutes for FDG and 18 minutes for AcAc, with doses of (338 $\pm$36) MBq and (190 $\pm$10) MBq, respectively. PET images were interpolated from (448, 448, 90) to (256, 256, 96) voxels, and CT images from (512, 512, 175) to the same resolution, with a voxel size of (1.17, 1.17, 2.75) mm, in order to make the data computationally manageable for training and to ensure consistent input dimensions across patients. All beds and other items that do not represent the patient were segmented out of the CT image. Fifteen patients of the dataset were selected to test data (7 FDG, 8 AcAc), 4 for the validation (2 FDG, 2 AcAc) and 140 for training. All data was clipped between -1000HU and 3000HU to mitigate the effects of metal artifacts on normalization range. Normalization was done to obtain values between -1 and 1 using equations \ref{eq:Nac} and \ref{eq:CT}:
\begin{equation}
    \widehat{NAC} = \left(\frac{NAC}{\max(NAC)} \times 2\right) - 1
    \label{eq:Nac}
\end{equation}
\vspace{-0.2cm}
\begin{equation}
    \widehat{CT} = \left(\frac{CT + 1000}{4000} \times 2\right) - 1
    \label{eq:CT}
\end{equation}

\subsection{Training}

\begin{figure}[H]
    \centering
    \includegraphics[width=1\linewidth]{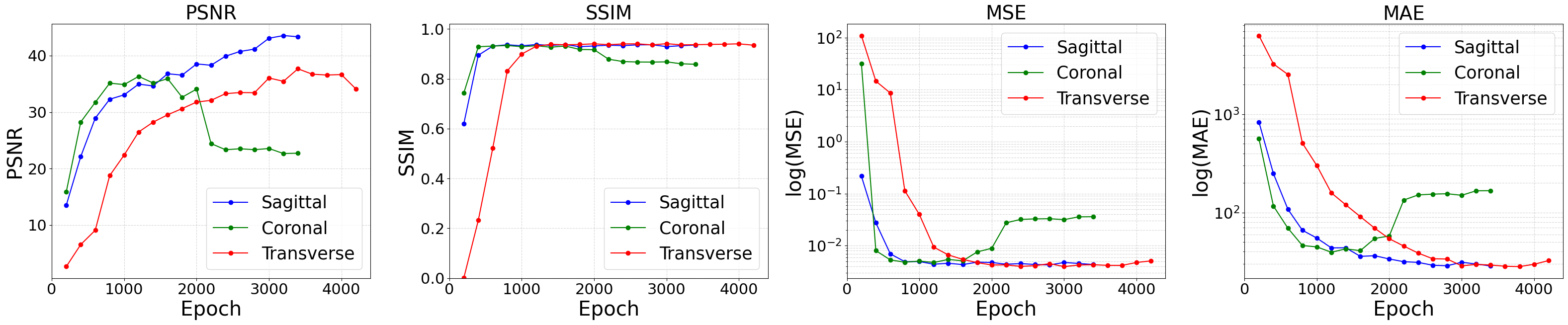}
    \caption{Progression of the validation metrics through the training.}
    \label{fig:validation_progression}
\end{figure}

The dataset was expanded by 60\% using data augmentation techniques including  $\pm$15° rotations,  $\pm$15\% zoom adjustments, and image flips. To augment images, we used a modified triangular distribution with uniform tails to bias sampling toward tissue-rich regions. This distribution assigns higher probabilities to central indices — where voids are less likely — and lower, uniform probabilities to outer regions. Specifically, the chosen index ranges were 0-83 (out of 96) in the transverse orientation, 30-235 in the coronal orientation, and 30-210 in the sagittal orientation (out of 256). To preserve anatomically plausible images, the augmentation strategies varied depending on orientation: extensive augmentation for transverse slices; only positive zoom and lateral flips for coronal slices; and exclusively positive zoom for sagittal slices. This approach yielded 21.560 transverse slices, 56,420 sagittal slices and 58,800 coronal slices for training. Noisy images were created using a cosine scheduler with 1,000 steps \cite{improved_DDPM}. The model utilized SNR-weighted MSE loss (min-SNR5) for training to improve training stability and sample quality \cite{snr}. During inference, EMA weights decay \cite{EMA} set to $\beta = 0.9995$ was used. The training process was initiated with 10 warm-up epochs, progressively raising the learning rate from 10$^{-6}$ to 10$^{-4}$ at each step with a cosine schedule. Subsequently, a second cosine scheduler reduced it from 10$^{-4}$ to 10$^{-6}$ over a 7 day period. Training was done on \textit{Tensorflow 2.15.1} with four \textit{Nvidia A100} 40GB GPUs using Tensorflow's mirrored strategy and mixed precision. A batch size of 80 per GPU for sagittal and coronal slices, and 40 for transverse slices was used. Training time depended on the orientation of the images, as some information was learned more quickly by the model. The transverse model took 3000 epochs to converge, while the sagittal model needed 3400 epochs and the coronal model 1200 epochs. \\
\\
To reduce overfitting, validation metrics were evaluated every 200 epochs using a single batch of images. Generating these validation images required 28 minutes for the transverse model and 22 minutes for the sagittal and coronal models. The progression of these metrics is shown in Figure \ref{fig:validation_progression}.\\
\\
Test samples were generated on an \textit{Nvidia A100} 40GB GPU, with batch sizes of 80, 120, and 120 for the transverse, sagittal, and coronal models, respectively. In the same order, generating the pCT image of one patient took 20, 22, and 22 minutes without mixed precision.

\subsection{Evaluation of PET quantification}

Reconstructed PET images were generated using Siemens' \textit{OPTOF} reconstruction algorithm. To obtain sharper corrected PET images, the automatic blur normally applied to the uncorrected PET image was removed, although this modification may slightly affect the results. PET images were reconstructed using two attenuation maps: the real CT available in the dataset (with dimensions \(256 \times 256 \times 96\)) and the corresponding pCT of identical dimensions, obtained after applying the majority voting. The PET image used for reconstruction is the original image that was not interpolated (\(448 \times 448 \times 90\)). The final attenuation corrected PET image therefore had the same dimension. The percentage difference between both reconstructions was computed using Equation~\ref{eq:PET_diff_precentage}:
\\
\begin{equation}
    Err = \frac{I_{\mathrm{pCT}} - I_{\mathrm{CT}}}{I_{\mathrm{CT}}} \times 100\%
    \label{eq:PET_diff_precentage}
\end{equation}
\\
where \(I_{\mathrm{CT}}\) denotes the voxel intensity of the PET image corrected using the real CT, and \(I_{\mathrm{pCT}}\) denotes the voxel intensity of the PET image corrected using the pCT. Ten Regions of Interest (ROIs) were extracted for both radiotracers using atlases provided by \textit{PMOD}. ROIs were selected to target regions in the brain that are studied for neurodegenerative diseases.

\section{Results}\label{sec3}

% DONE Mettre tableau avec les métriques de l'ensemble des erreurs pour chaque patients (box plot aussi possible) . 

%Dire que PSNR et SSIM sont avec un range de -1 à 1 et mentionner quelle métrique a été calculée avec les données norm vs not norm.

%Colormaps

%Tableau avec tt les ROI

% DONE Comparaison des CT générés vs real CT en métriques

%Voting image

%Variance de la génération d'images (génère plusieurs fois le même patient et voit la différence entre les générations). Higher STD = lower confidence. 

% DONE Mettre le training progress pour montrer l'overfit pour chaque modèle ???

% # Compute pixel-wise statistics
% mean_prediction = np.mean(samples, axis=0)
% std_prediction = np.std(samples, axis=0)
% confidence = 1 / (1 + std_prediction)  # Higher std = lower confidence

\begin{figure}[H]
    \centering
    \includegraphics[width=0.8\linewidth]{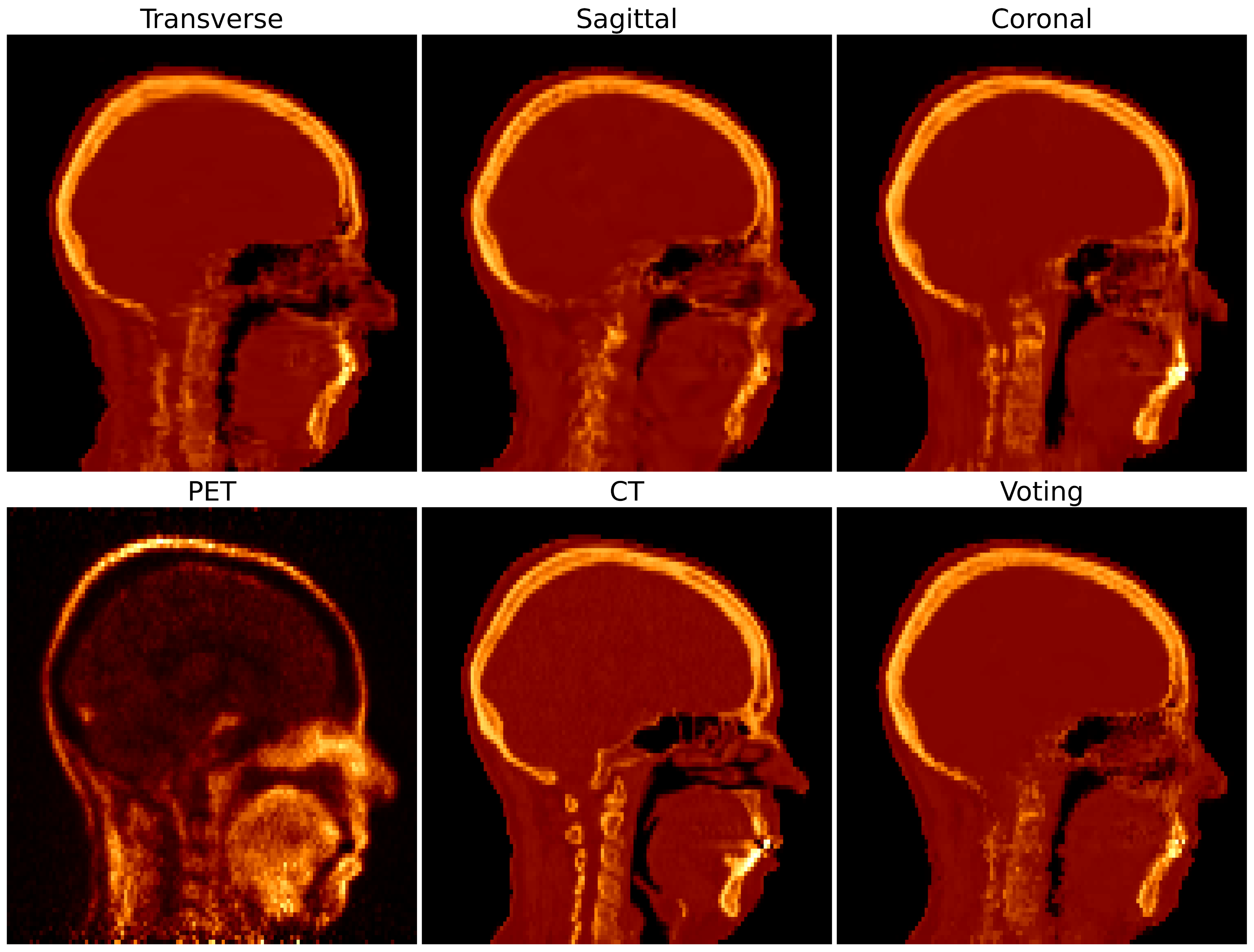}
    \caption{Sagittal views of the  pCTs generated by the Transverse, Coronal and Sagittal models, respectively, \textit{(top)} compared to the NAC PET image \textit{(bottom left)}, ground truth CT \textit{(bottom middle)} and  pCT after voting \textit{(bottom right)} using the AcAc tracer in Patient 34028.}
    \label{fig:side_by_side_CT}
\end{figure}

\begin{figure}
    \centering
    \includegraphics[width=1.1\linewidth]{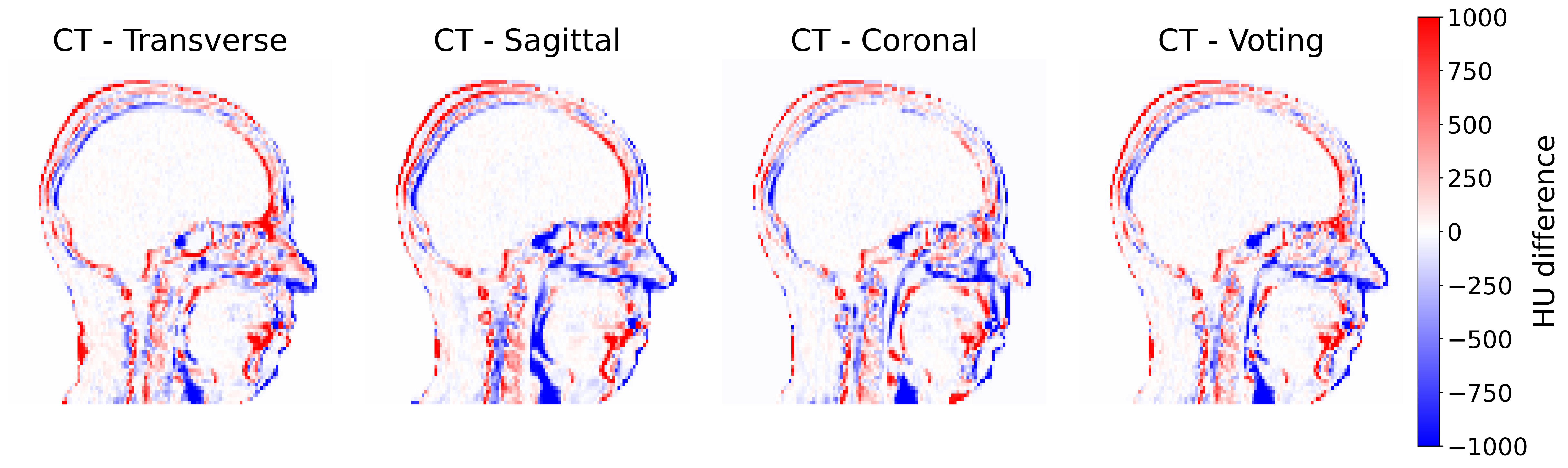}
    \caption{Difference map of patient 34028 between the CT and pCT for each model's prediction and voting.}
    \label{fig:side_comparison}
\end{figure}

The generated pCTs are visually compared with the CT and NAC PET AcAc images used to generate the pCTs in Figure \ref{fig:side_by_side_CT}. Each model's image shows a great generation of the skull and the brain with very little visual difference from the CT. All three models struggle in the sinus region, where predictions of cavities filled with air and cervical bones differ substantially. The transverse model shows lack of consistency in the airway and the back of the neck's shape is inconsistent. The sagittal model shows a smoother back neck generation, but fails to adequately generate the airway and has difficulties generating the cervical bones. The coronal model's image provides a clear and consistent airway, but it appears completely closed at the bottom of the image. The voting image shows a well shaped neck, but the issues from the airway persist as the majority of the models failed to correctly generate it. Patient 34028 is further analyzed in a difference map in Figure \ref{fig:side_comparison}. Results of the average metrics of all 15 pCT from the test data set are quantitatively compared to their corresponding CT in Table \ref{tab:metrics}.

\vspace{-0.3cm}

\vspace{-0.6cm}

\begin{figure}[H]
    \centering
    \includegraphics[width=0.7\linewidth]{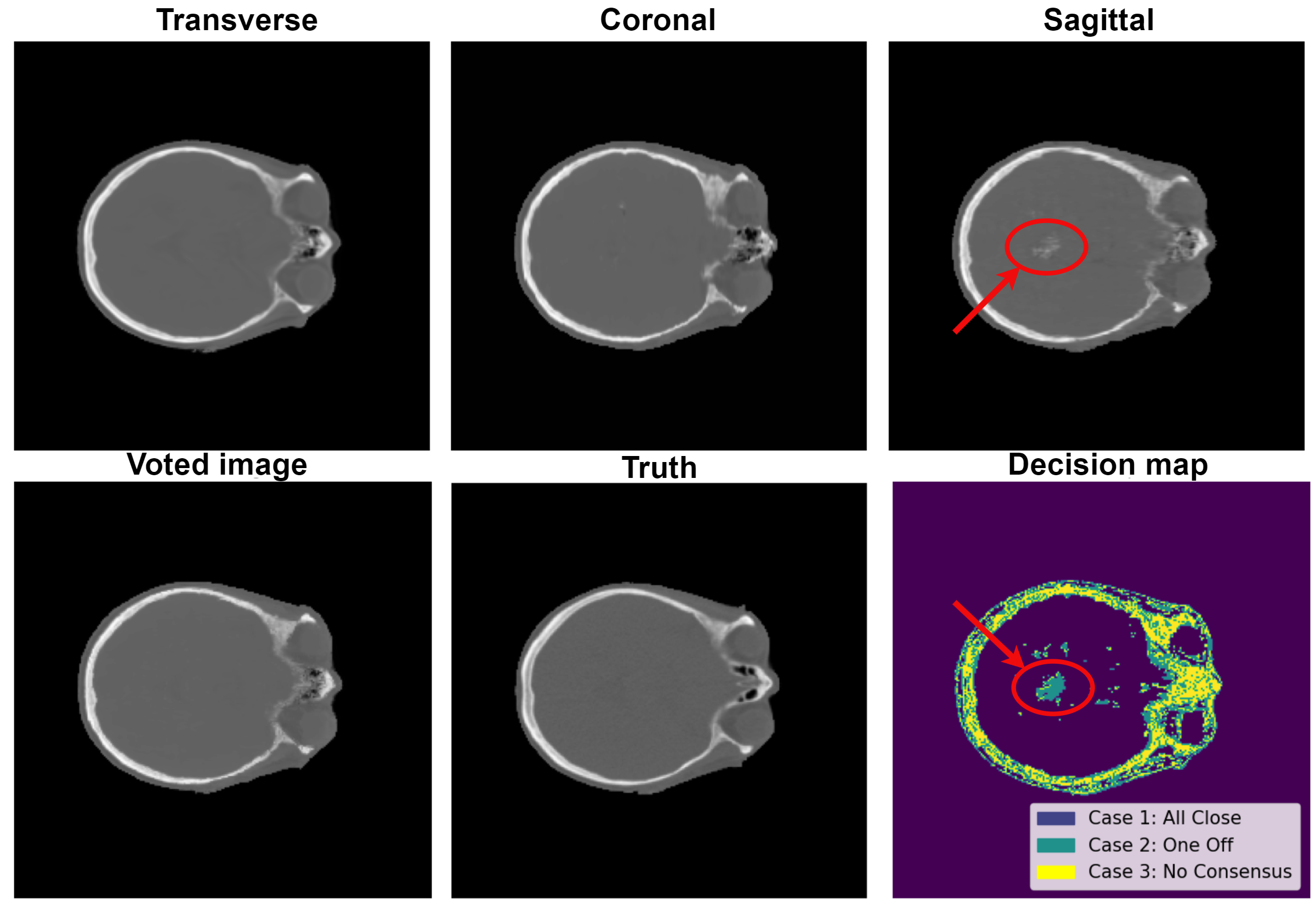}
    \caption{Majority voting removing artifacts present in the sagittal model's image of patient 34024. Decision map: [Dark blue] All values from different images within threshold, [Green] One value above threshold, [Yellow] Two or more values above threshold.}
    \label{fig:voting_artifact_removal}
\end{figure}

\begin{table}[!h]
\centering
\caption{Average metrics and standard error (STD) of the pCT for the 15 patients of the test data set compared to the corresponding CT for each model.}
\renewcommand{\arraystretch}{1.5}
\begin{tabular}{|p{2.5cm}|>{\centering\arraybackslash}p{2.6cm}|>{\centering\arraybackslash}p{2.6cm}|>{\centering\arraybackslash}p{2.6cm}|>{\centering\arraybackslash}p{2.6cm}|}
\hline
\textbf{Method} & \textbf{Transverse} & \textbf{Coronal} & \textbf{Sagittal} & \textbf{Voting} \\
\hline
PSNR (dB) ($\uparrow$) & 29.9 $\pm$ 2.3 & 29.4 $\pm$ 2.1 & 29.8 $\pm$ \textbf{2.0} & \textbf{30.1} $\pm$ 2.2 \\
\hline
SSIM ($\uparrow$) & 0.924 $\pm$ 0.030 & 0.916 $\pm$ 0.032 & 0.919 $\pm$ 0.030 & \textbf{0.925 $\pm$ 0.029} \\
\hline
MAE (HU) ($\downarrow$) & 31.7 $\pm$ 11.9 & 40.8 $\pm$ 10.9 & \textbf{30.2} $\pm$ 10.8 & 32.1 $\pm$ \textbf{10.4} \\
\hline

RMSE (HU) ($\downarrow$) & 131.8 $\pm$ 33.9 & 138.7 $\pm$ 31.7 & 133.2 $\pm$ \textbf{29.5} & \textbf{128.5} $\pm$ 31.0 \\
\hline
Avg Error (HU) ($\downarrow$) & -0.15 $\pm$ 6.83 & -12.15 $\pm$ 3.86 & \textbf{0.34} $\pm$ 4.02 & -3.80 $\pm$ \textbf{3.76} \\
\hline
\end{tabular}
\begin{tablenotes}
\footnotesize
\item[*] $\uparrow$: higher is better, $\downarrow$: lower is better
\end{tablenotes}
\label{tab:metrics}
\end{table}
\vspace{-0.3cm}
\vspace{-0.3cm}
These metrics show that voting with a very loose 50 HU threshold helps the pCT to more closely match the CT. This is shown by a higher PSNR, SSIM, lower MAE, lower RMSE and lower STD of the average error. All models show similar STD on all metrics, showing that predictions are consistent. Without the voting method, all metrics, showing that predictions are consistent. Without the voting method, all three models show very similar results in similarity metrics (SSIM and PSNR). The coronal model has higher error metrics with the average error pointing to an average under estimation of the values by -12 HU.\\\\
Figure \ref{fig:voting_artifact_removal} shows that majority voting can remove artifacts generated inside the brain by the sagittal model along with other small artifacts that cannot be seen visually.

\vspace{-0.4cm}
\vspace{-0.5cm}

% Reconstructions
Preliminary reconstructions were performed on four patients injected with an \textsuperscript{18}F-FDG tracer, and the activity within each ROIs was calculated. The results are presented in Table~\ref{tab:ROI_regions}. The mean error across all 116 ROIs provided by the atlas, which covers most of the brain, was (1.46 $\pm$ 0.68)\%.\\\\

\begin{table}[h!]
\centering
\caption{ROI difference between CTAC and pCTAC PET images for 4 FDG patients.}
\begin{tabular}{lcc}
\hline
\textbf{Region} & \textbf{Mean $\pm$ STD [\%]} & \textbf{Volume (ccm)} \\
\hline
Amygdala          & $1.04 \pm 0.85$ & 7.49 \\
Caudate           & $2.06 \pm 0.93$ & 31.30 \\
Cingulum Post    & $1.73 \pm 0.92$ & 12.77 \\
Frontal Mid      & $1.99 \pm 1.45$ & 159.47 \\
Frontal Sup      & $2.39 \pm 1.69$ & 122.48 \\
Hippocampus       & $1.08 \pm 0.81$ & 30.05 \\
Parahippocampus   & $0.90 \pm 1.02$ & 33.76 \\
Precuneus         & $2.13 \pm 1.61$ & 108.69 \\
Temporal Mid     & $0.75 \pm 0.45$ & 149.62 \\
Temporal Sup     & $1.02 \pm 0.52$ & 86.99 \\
\hline
\end{tabular}
\label{tab:ROI_regions}
\end{table}

\begin{figure}[H]
    \centering
    \includegraphics[width=0.7\linewidth]{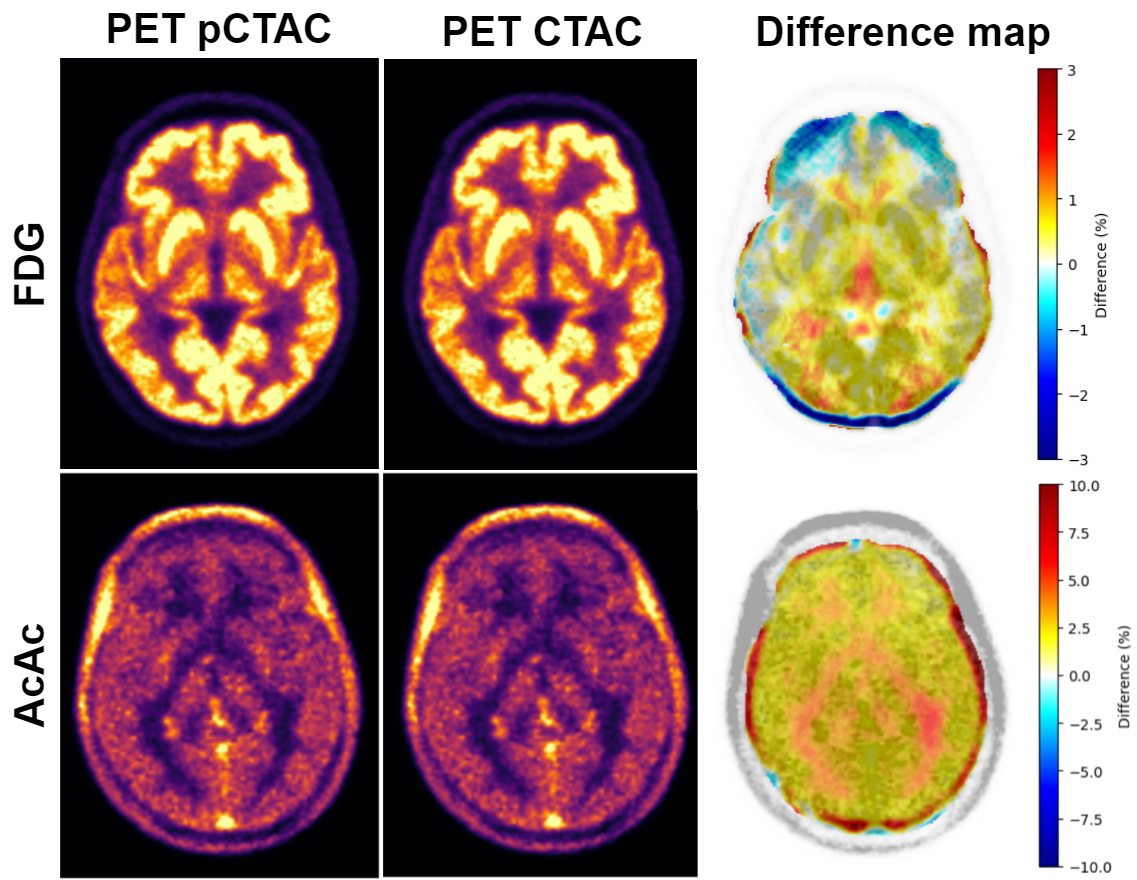}
    \caption{Visual comparison of reconstructed PET images using a CT vs pCT attenuation map of patient 33609 FDG.}
    \label{fig:PET_corrected_comp}
\end{figure}

Figure~\ref{fig:PET_corrected_comp} shows a visual comparison between a PET images reconstructed using a pCT and a real CT. Visually, for both tracers, the two images show close to no difference. The corresponding difference map supports this observation, showing an average pixel error of (0.58 $\pm$ 0.50)\% for the FDG and (2.92 $\pm$ 1.68)\% for the AcAc.

\vspace{-0.4cm}
\vspace{-0.4cm}

\begin{figure}[H]
    \centering
    \includegraphics[width=1\linewidth]{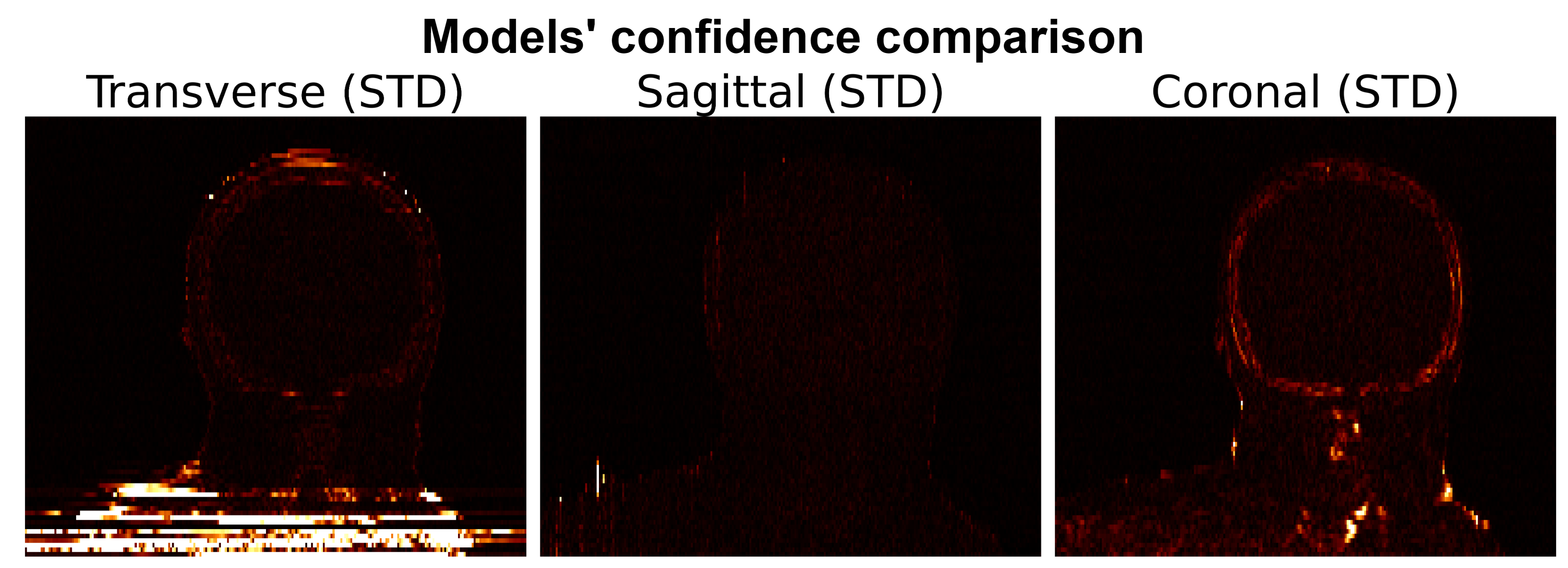}
    \caption{STD variation by model for three predictions of the same patient (33011 AcAc). Displayed in the coronal view. Range of the displayed STD set from 0 to 5 HU.}
    \label{fig:confidence_image}
\end{figure}

%Model confidence on CT images
To assess the confidence of the three coronal, sagittal and transverse models, the same test dataset was generated three times per model. A strong confidence model should demonstrate minimal variability in its predictions. Table \ref{tab:confidence_analysis} reports similarity scores (PSNR and SSIM) that are extremely high across all models, showing near identical images. Overall, all scores indicate strong, confident models. The biggest variations occur towards the shoulders of the patient (if they are visible) as can be seen in Figure \ref{fig:confidence_image}.

\vspace{-0.3cm}
\begin{table}[htbp]
\centering
\caption{Model's confidence analysis metrics for 3 generation of the same image for 15 patients.}
\renewcommand{\arraystretch}{1.5} % Increase row height
\begin{tabular}{|l|c|c|c|}
\hline
\textbf{Metric} & \textbf{Coronal} & \textbf{Transverse} & \textbf{Sagittal} \\[0.5ex]
\hline
Pairwise PSNR (dB) ($\uparrow$) & $61.2 \pm 4.8$ & $68.9 \pm 11.3$ & $63.1 \pm 8.5$ \\[0.5ex]
\hline
Pairwise SSIM ($\uparrow$) & $0.9998 \pm 0.0006$ & $0.9998 \pm 0.0008$ & $0.9998 \pm 0.0004$ \\[0.5ex]
\hline
Mean Coefficient of Variation (CV) ($\downarrow$) & $0.124 \pm 0.464$ & $0.091 \pm 0.336$ & $0.345 \pm 0.969$ \\[0.5ex]
\hline
Mean voxel-wise std ($\downarrow$) & $0.22 \pm 0.23$ & $0.27 \pm 0.60$ & $0.10 \pm 0.05$ \\[0.5ex]
\hline
95th percentile voxel std ($\downarrow$) & $0.60 \pm 0.78$ & $0.27 \pm 0.17$ & $0.17 \pm 0.03$ \\[0.5ex]
\hline
\end{tabular}
\label{tab:confidence_analysis}
\end{table}
\vspace{-0.3cm}

%Image montrant les cas difficiles de confiance
\begin{figure}[H]
    \centering
    \includegraphics[width=1\linewidth]{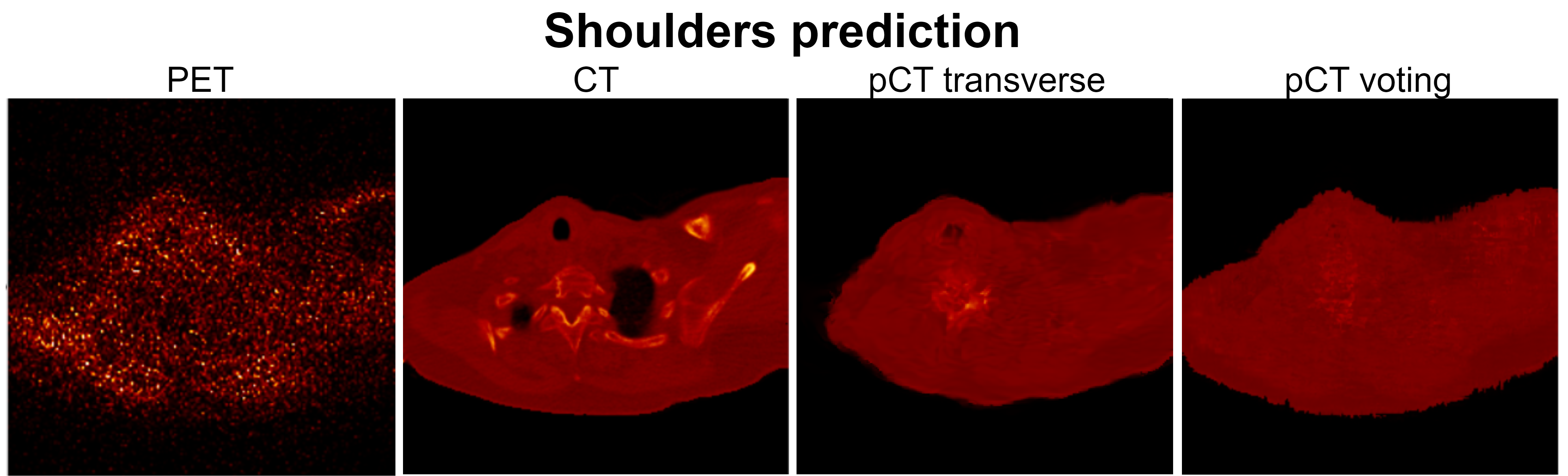}
    \caption{Generation of a transverse and voting pCT image from a PET image at shoulders level compared to the CT for patient 33011 AcAc.}
    \label{fig:confidence_transverse}
\end{figure}

\vspace{-0.4cm}
The transverse model shows variation across its three predictions at the shoulders, with standard deviations reaching up to 500 HU for the same input image. In contrast, the sagittal model's predictions are almost identical, while the coronal model only shows minor variation, mostly in the bone regions. An example of a transverse PET slice that resulted in a high-variability pCT prediction of the shoulders is shown in Figure \ref{fig:confidence_transverse}. It appears very noisy and with little to no recognizable structures. Applying majority voting reduces this variability and improves the prediction.

\begin{figure}[H]
    \centering
    \includegraphics[width=0.5\linewidth]{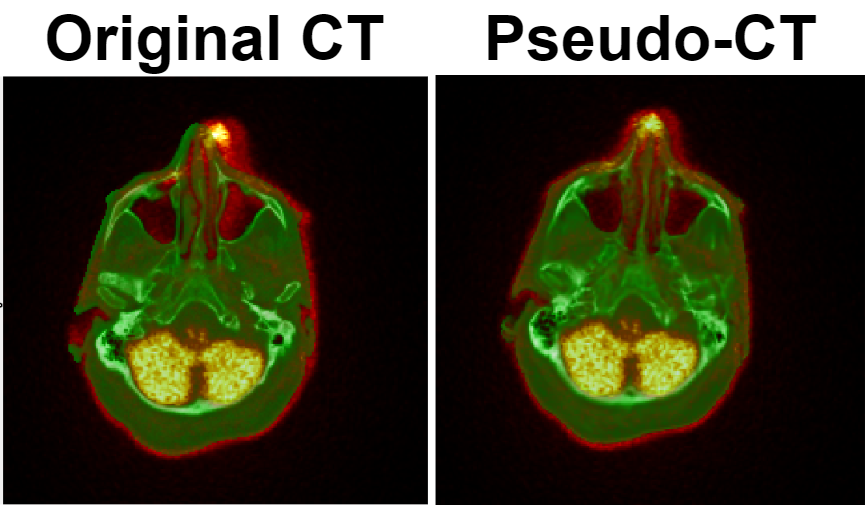}
    \caption{Overlapping of a CT image (green) with a PET image (red) of patient 34113. [Left] Overlapping with the CT; [Right] Overlapping with the  pCT.}
    \label{fig:movement}
\end{figure}

Figure \ref{fig:movement} shows a patient in the test dataset that moved between the CT and PET acquisitions. The patient's head tilted to the right after the CT scan (in green) was completed, as evidenced by the noses not being aligned with the PET image (in red) on the left image. Looking closely at the nose of the PET image, the patient also moved during the PET acquisition as there is some ghosting to the right of the hotspot of the tip of the nose. The image generated by the model, on the right, is well aligned with the PET image as it only relies on the PET image to generate the pCT.
\vspace{-0.4cm}
\vspace{-0.4cm}

\begin{figure}[H]
    \centering
    \includegraphics[width=1\linewidth]{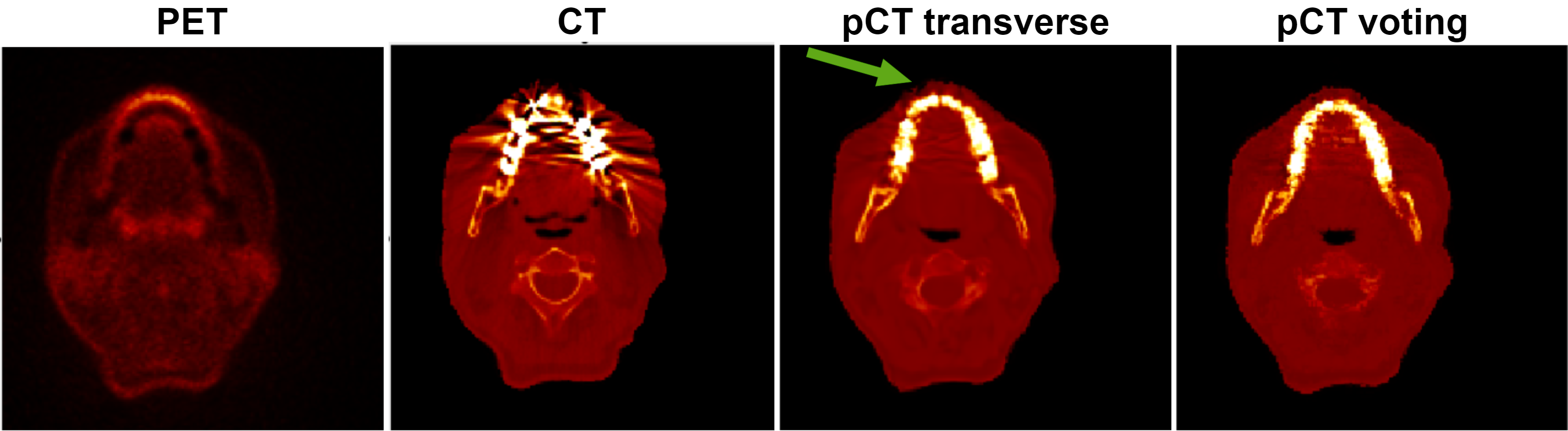}
    \caption{CT and pCTs of Patient 34024 FDG with metal dental fillings.}
    \label{fig:artefact_teeth}
\end{figure}

Figure \ref{fig:artefact_teeth} shows a CT image with metal streak artifacts from dental implants, which produce over- and under-estimations across the image. The pCT generated by the transverse model substantially reduces these artifacts, but has parts of the chin missing (indicated by the green arrow). The use of majority voting corrects this error.
\vspace{-0.4cm}

\section{Discussion}
Only four reconstructions with a single radiotracer were shown quantitatively in this current article because it is still a work in progress. All reconstructions with both tracers will be included in the final version of the article.\\\\
The bed and other external objects (e.g., oxygen tubes) were segmented out of the CT images so the model could focus only on relevant anatomical information. This segmentation was not applied to the PET images, since doing so would create artificial discontinuities (regions set to zero) that would disrupt the natural noise pattern of PET images. These discrepancies may increase the risk of the model learning irrelevant artifacts. For instance, these inconsistencies between the PET and CT images might introduce attenuation patterns that do not appear in the corresponding CT image. Although training without segmentation of external objects was not tested, the results obtained using segmented CT images suggest that the model effectively learns to disregard these inconsistencies. Instead, the model focuses on the relevant element of the image: the patient's head.\\\\
%
%
%Voting threshold value
The choice to use a threshold value for voting of 50 HU is meant to increase the accuracy for the brain area. Most commonly found matter inside this region ranges approximately from 0 (cerebrospinal fluid) to 45 HU (gray matter). Predicted values outside this range can be considered aberration and should not be included in the final image. Voting has little effect on regions with higher values, like bones, where simple fluctuations in these regions can be more than the defined threshold, leading to the model simply averaging these areas in most cases.\\\\
%
%temps Convergence diff 
Training took a different number of epochs for each model, as seen in Figure \ref{fig:validation_progression}, with the most noticeable difference coming from the coronal model that took about 2000 less epochs to converge compared to the sagittal and transverse models. There is no clear explanation for this behaviour. All models had the exact same training parameters and seeds, the only difference being the orientation of the data fed to the models. One argument can be that data in the coronal view is more repetitive: A human head is longer (coronal) than it is large (sagittal), resulting in more anatomical information seen per epoch. This, added with the fact that the coronal shape is quite repetitive between image slices could be the reason of this faster training yielding similar results to the other models. \\\\
%
%Ressources coronal
This is an important consideration when looking at the computational resources required to train such models. In our case, the training time to convergence (no validation calculations, purely time per epoch time number of epoch) was 56.6 hours for the coronal model, 112.5 hours for the transverse model, and 153 hours for the sagittal model. This shows a large disparity in training costs. With limited resources, training a single coronal model is far more efficient, as it achieves comparable results while requiring only a third of the computation time. Figure \ref{fig:side_by_side_CT} reveals that diffusion models have their greatest inaccuracies in the nose and sinuses. These regions contain air and thin, low-uptake tissue with minimal detail in the input PET image, making these features difficult to generate accurately. This also applies to cervical bones that show low uptake for both FDG and AcAc tracers, making it also difficult to generate the corresponding image. These regions are also unique to each patient, where everyone has varying sinus \cite{sinus_variation} and cervical bone \cite{vertebre_cervical} shapes. The learning of these features therefore becomes very challenging for the model that only has access to the PET image. The voting image contains features of all three images, but without any significant visual changes. Images from both radiotracers yielded similar results.\\\\
%
%
%Voting justification
The voting method offers many advantages, such as improving most metrics and effectively removing artifacts produced by individual models, as shown in Figure \ref{fig:voting_artifact_removal}. We compared it with two alternatives: simple averaging and taking the median value. While both averaging and median methods achieved slightly better quantitative metrics, they come with drawbacks. Averaging provides good visual results, but does not remove artifacts as it only attenuates them, not solving the artifact problem. The median method can remove artifacts, but it discards valuable information from other models, leading to visual differences in some regions. Voting combines the strengths of both approaches, producing high-quality images that are both visually convincing and artifact-free.\\\\
%
%Reconstructions
Typically, in clinical images, all uncorrected PET images are systematically blurred using a heavy gaussian filter before applying the attenuation map. This produces the blurred PET images commonly used in the clinic, but for demonstration purposes we decided to skip this filtering step. Gaussian smoothing reduces high-frequency differences between the images, which are typically the main contributors to error metrics. As a result, omitting the filter tends to worsen these metrics. Nevertheless, it provides sharper reconstructed PET images, which is more valuable for us, as it allows precise identification of where the errors occur inside the images.\\\\
%
%Metriques tableau
Visually, looking at Figure \ref{fig:side_comparison}, all the models show a strong performance inside the brain region where maximal errors can reach up to 50 to 100 HU. Most errors come from the bones that have a lower uptake value and, therefore have less information for a prediction. Table \ref{tab:metrics} shows the coronal model is slightly worse in terms of errors. This can be attributed to the smaller training time resulting in a shorter convergence point for the model. Indeed, the validation metrics were taken every 200 epochs and looking at Figure \ref{fig:validation_progression}, there is a steep drop in metrics quality between epoch 1200 (best epoch) and 1400. This shows that the true convergence point may lie at epoch 1200  $\pm$ 200. This point also applies to the other two models, but the drop in metrics quality before and after convergence is more subtle than for the coronal model.\\\\
%
%
%Validation metrics
This issue could have been avoided with more frequent validation, but we initially did not expect the coronal model to converge so quickly. However, increasing validation frequency comes at a cost: generating a single batch of validation images takes 22 or 28 minutes, depending on the model. For context, the transverse model required 9.8 hours just to generate validation samples over its 4200 training epochs. More frequent validation steps could be used, but would significantly extend total training time and resource usage.\\\\
Even with a threshold set to 50 HU (a value determined based on the brain's values), voting helps the overall image scores to an extent where is outperforms the other models in PSNR, SSIM, MAE STD, RMSE and average STD error. Although it was discussed that voting might have minimal impact on bones or structures with higher HU values, two models can still agree on the value while the third model provides a value above the threshold. This scenario is less likely due to the amplitude of the values varying much, as mentioned, but can still help generating an improved image - In the worst case, it will simply average all values, so if one model really messes up a prediction, the other two models will attenuate the repercussions of this wrong prediction. \\\\ 
%
%
%PET quantifications VISUAL
%Results in Table \ref{tab:ROI_regions} show very small difference between the PET images reconstructed with a CT versus a pCT for both FDG and AcAc tracers. AcAc tracer shows a higher average error, because it has lower activity inside the brain. This can be verified looking at table \ref{XXX} where the average concentration is more than 2 times lower than FDG. Therefore, the statistical error is increased and gives these higher error numbers, even though the images look visually like the same quality. In Figure \ref{fig:PET_corrected_comp}, it can be seen that the main errors occur at bone/air interfaces where significant density changes take place. Precisely identifying where these structures end can be difficult due to partial volume effects or patient movement during PET imaging, which explains the pseudo-CT's struggles in these areas. This shows, as the hippocampus (1.08 $\pm$ 0.81) and parahippocampus (0.90 $\pm$ 1.02) that are deep inside the brain have the lowest error, while the frontal mid (1.99 $\pm$ 1.45) and sup (2.39 $\pm$ 1.69) are on the edge of the brain and show higher error.\\\\
Results in Table \ref{tab:ROI_regions} show very small difference between the PET images reconstructed with a CT versus a pCT for FDG. AcAc tracer shows a higher average error at figure \ref{fig:PET_corrected_comp}, because it has lower activity inside the brain, therefore, the statistical error is increased and gives these higher error numbers, even though the images look visually like the same quality. In Figure \ref{fig:PET_corrected_comp}, it can be seen that the main errors occur at bone/air interfaces where significant density changes take place. Precisely identifying where these structures end can be difficult due to partial volume effects or patient movement during PET imaging, which explains the pseudo-CT's struggles in these areas. This shows, as the hippocampus (1.08 $\pm$ 0.81) and parahippocampus (0.90 $\pm$ 1.02) that are deep inside the brain have the lowest error, while the frontal mid (1.99 $\pm$ 1.45) and sup (2.39 $\pm$ 1.69) are on the edge of the brain and show higher error.\\\\
%
%Confiance du model
Due to the stochastic nature of DDPM models, each generated prediction differs slightly, because random noise is used for each image generation. Verifying whether the model can produce consistent outputs from different noise initializations is essential, as it reflects the model's confidence. An uncertain model will generate varying images with the same input condition, but a confident model will consistently produce the same result. As shown in Figure \ref{fig:confidence_image}, the transverse model struggles to consistently reconstruct the shoulders. This happens because the PET  image used to generate the transverse pCT, seen in Figure \ref{fig:confidence_transverse}, is very noisy: the shoulders show low tracer uptake and are further affected by edge-of-field noise amplification. As a result, the image contains very little usable information for the model to generate a patient-specific reconstruction. In contrast, the coronal and sagittal models are less affected, since edge-of-field noise amplification primarily occurs in the axial FOV. This would suggest that only the coronal or sagittal model should be used to generate the bottom transverse slices as they generate more confident results.\\\\%PAs certain de cette partie...
%
%
%Patients qui bougent
The left image in Figure \ref{fig:movement} shows a patient who moved between the CT and PET scans. This is a common clinical challenge where the CT is acquired in just a few seconds, while the PET scan lasts several minutes. Patient movement in between can compromise attenuation correction, potentially leading to poorer image quality and misdiagnosis. Generating a pCT image from PET alone completely avoids this issue, as the pCT is always perfectly aligned with the PET, as illustrated by the right image in Figure \ref{fig:movement}. We could even argue that in case the case where a patient moved, the pCT is more realistic than the CT for attenuation correction.\\\\
%
%
%Mis alignement
Even small patient movements between scans can complicate model training. In most cases, the CT is well aligned with the PET, but when misalignment occurs, the model is penalized for producing a pCT image aligned with the PET image, which can degrade training performance. In addition, such misalignment affects evaluation metrics, where in Table \ref{tab:metrics}, the CT-PET mismatch is not accounted for, meaning the reported metrics are pessimistic and would improve if all data were perfectly aligned. Furthermore, all metrics except SSIM are evaluated pixel by pixel meaning that a slight movement will impact negatively the metrics. We assumed that majority of patients inside the dataset moved a negligible amount and we can confirm this because all pCT predictions look correctly aligned with the PET.\\\\
%
%
%Metal artifacts
As shown in Figure \ref{fig:artefact_teeth}, some patients in the dataset had dental fillings, which pro- duced metal streak artifacts \cite{artefact_ct}. These artifacts create over- and underestimations in the transverse view and can mislead the model, which may attempt to reproduce them. Fortunately, because many patients in the training dataset did not show such artifacts, the model can learn to generate artifact-free images. This is something conventional CT cannot reliably achieve, highlighting a potential advantage of this AI-based approach. Furthermore, training on a dataset that entirely excludes patients with dental fillings might prevent the model from ever generating artifacts in such cases, simply because it never learned them. In the current setup, the model is penalized whenever it generates a clean image while the ground truth still contains artifacts.\\\\ 
%
%
%Limitations
Many limitations were encountered in this project. First, the dataset didn't contain any particular conditions or pathologies. An image could have something the model has never seen such as a large tumors, metal plates, missing skull parts, neuro-degenerative diseases, etc. It is unknown how the model would react in these scenarios and to eventually deploy this model, these concerns must be addressed.\\\\ 
%
%Limitations of majority voting
Although majority voting shows great potential, its application in this work was rather simplistic. We acknowledge that more complex implementation may produce better results with the three images provided by the models.\\\\
Additionally, data obtained from different PET scanners often show considerable differences in resolution, noise levels, contrast, and overall image texture. Since neural networks are highly sensitive to such variations, a model trained on data from a single scanner may capture some scanner-specific characteristics instead of true anatomical or physiological information. When this model is tested on images from another scanner, the mismatch can result in degraded performance, reduced accuracy, visible artifacts, or limited generalization ability. \\\\
Furthermore, patient movement introduces an additional challenge where it causes variability in the ground truth that the model is trained to learn. One possible solu- tion is to use an image registration algorithm to align all scans, but this approach can also introduce bias into the dataset from the algorithm itself.\\\\

\section{Conclusion}

This work introduced a new approach using Multiview Denoising Diffusion Probabilistic Models to generate convincing  pCT images for attenuation correction of PET images . The use of majority voting with a simple threshold of 50 HU improves almost all metrics of the pCT while removing artifacts that may have been generated by one of the transverse, coronal or sagittal models.\\\\
With majority voting, PSNR improved from (29.93 $\pm$ 2.26) dB from the best model (transverse) to (30.13 $\pm$ 2.16) dB for the voting and RMSE was reduced from (131.79 $\pm$ 33.93) HU (transverse) to (128.46 $\pm$ 31.00) HU. The nasal and sinus areas present difficulties for the model, as PET images provide limited information about these anatomically complex regions, but the model showed strong performance in the brain region.\\\\
We demonstrated that this approach can successfully generate accurate, well aligned and artifact free deep learning-based pCT images from non-attenuation-corrected PET data, allowing attenuation correction of brain PET scans acquired with two different radiotracers showing distinct uptake distributions. By eliminating the need for additional imaging modalities and avoiding extra radiation exposure, this method has the potential to make brain imagingsafer, more cost-effective and free from misalignment errors.

\section*{Declarations}
\textbf{Ethics approval: }  
\textbf{Data availability: } The dataset used and analyzed in this study is available from the corresponding author on reasonable request.\\
\textbf{Funding: } 
\textbf{Competing interests: } RL is both with the Université de Sherbrooke and a co-founder and Chief Scientific Officer of IR\&T Inc., which funded part of this work through the Acuity-QC Consortium. SC consults for and has received research funding and research materials from Nestlé Health Science. He also consults for Cerecin. Other authors have no competing interests to declare relevant to the content of this article.\\
\textbf{Author contributions: } \\ASG: Conceptualization, DDPM implementation, Software development, Data analysis, Writing original draft, review \& editing; \\
MT: Supervision, Data analysis, Review \& editing; \\
CT, ÉA: Technical assistance, Review \& editing; \\
ÉC, SC: Data procurement, Review \& editing; \\
RL: Funding acquisition, Supervision, Review \& editing; \\
JBM: Funding acquisition, Conceptualization, Supervision, Review \& editing.  \\
All authors gave their approval for the final version of the manuscript.\\
\textbf{Acknowledgements: }Not applicable.

\newpage
\bibliography{sn-bibliography}

\end{document}